\documentclass[12pt]{article}
\usepackage{graphicx}
\usepackage[cp1251]{inputenc}
\usepackage{rotating}
\begin{document}
 \noindent {\footnotesize\it Astronomy Reports, 2018, Vol. 62, No 9, pp. 557--566}
 \newcommand{\dif}{\textrm{d}}

 \noindent
 \begin{tabular}{llllllllllllllllllllllllllllllllllllllllllllll}
 & & & & & & & & & & & & & & & & & & & & & & & & & & & & & & & & & & & & & \\\hline\hline
 \end{tabular}

 \vskip 0.5cm
  \centerline{\bf\large A Search for a Globular Cluster whose Passage Through}
  \centerline{\bf\large the Galactic Disk Could Induce the Formation of the Gould Belt}
 \bigskip
 \bigskip
  \centerline
 {
 V.V. Bobylev and A.T. Bajkova
 }
 \bigskip
 \centerline{\small \it
 Central (Pulkovo) Astronomical Observatory, Russian Academy of Sciences,}
 \centerline{\small \it Pulkovskoe shosse 65, St. Petersburg, 196140 Russia}
 \bigskip
 \bigskip
 \bigskip

 {
{\bf Abstract}---The distribution of sites where globular clusters
have crossed the Galactic disk during the last 100 million years
has been analyzed using the most recent kinematic data for 133
globular clusters (GCs). Three GCs (NGC 6341, NGC 7078, and
$\omega$ Cen) whose distances between the positions where they
crossed the Galactic disk and trajectories of the Gould Belt are
less than 20\% of their heliocentric distances at the crossing
time (82, 98, and 96 million years ago, respectively) have been
identified. For each of the clusters, this was their next to last,
rather than their last, crossing of the Galactic disk. The passage
of any one of these three GCs through the disk could potentially
have initiated the formation of the Gould Belt.
  }

\medskip DOI: 10.1134/S1063772918090020

 \section{INTRODUCTION}
A giant stellar--gas complex, known as the Gould Belt, is located
near the Sun. This structure includes several OB and T
associations, dozens of open star clusters, numerous single young
stars, and molecular and dust clouds. According to modern
estimates, the Gould Belt is a fairly flat system, with semi-axes
of $\approx350\times250\times50$ pc and an inclination to the
Galactic plane of about 20$^\circ$. The center of the system is
about 150 pc from the Sun, in the second Galactic quadrant. The
stars forming the Gould Belt have ages of less than 60 million
years. The total mass of the complex is approximately
$1.5\times10^6 M_\odot.$ For a detailed description of the
properties of this structure, see [1--6].

Quite a number of models have been suggested to explain the origin
and evolution of the Gould Belt. According to Blaauw [7], this
structure could have arisen as a result of the expansion of gas
accelerated to high velocities following the explosion of a
hypernova. This idea was further developed by Olano [8] in the
framework of a gas-dynamical model. Lindblad [9] suggested a
purely descriptive kinematic linear model for the intrinsic
differential rotation and expansion of the Gould Belt that agree
fairly well with observations. Bobylev [10, 11] further developed
this approach for the non-linear case. Olano [12] considered an
evolutionary model for a super-cluster, in which clusters such as
the Hyades, the Pleiades, Coma Berenices, and the Sirius cluster
were fragments of a former single complex, and the formation of
the Gould Belt was a result of the contraction of the parent
cloud's central parts. None of these models are able to explain
the observed inclination of the Gould Belt to the Galactic plane,
while the three-dimensional model of Perrot and Grenier [13]
describes the evolution of the inclination of the Gould Belt
during a time interval about 50 million years into the past.

Another series of models consider a massive body falling onto the
Galactic plane. L\'epine and Duvert [14] suggested that many
molecular-cloud complexes near the Sun could have been formed due
to collisions between high-velocity clouds and the Galactic disk.
In this model, the Gould Belt could have formed purely by chance.

Comer\'on and Torra [15, 16] performed numerical simulations of
the passage of a high-mass ($3.3\times10^6 M_\odot$),
high-velocity cloud through the disk, and found that star
formation would be most effective in the case of an oblique impact
of the falling body. Levy [17] demonstrated that a shock could
arise in the gas if a globular cluster (GC) crossed the Galactic
disk. Finally, Bekki [18] demonstrated that the Gould Belt could
have been formed some 30 million years ago as a result of a
high-velocity, oblique impact between a clump of dark matter with
a mass of approximately $3\times10^7 M_\odot$ and a gas cloud with
a mass of about $10^6 M_\odot.$ He found that the dynamical action
of the falling body transformed an ellipsoidal cloud into a
stellar ring structure after it has induced star formation in the
cloud. His computations also showed that the resulting stellar
structure should have an appreciable inclination to the Galactic
disk, since the cloud becomes elongated along the orbit of the
massive clump.

The masses of several GCs in the Galaxy reach a few times $10^6
M_\odot.$ It is quite possible that such a GC induced the
formation of the Gould Belt. It is of interest to search for a
particular GC capable of doing this. High-accuracy measurements of
the proper motions, radial velocities, and distances are now
available for many GCs, making it possible to derive their
Galactic orbits over long time intervals.

The aim of this paper is to study the Galactic orbits of known GCs
in order to search for a suitable candidate whose passage through
the Galactic disk could have triggered the formation of the Gould
Belt.

 \section{DATA}
We based our study on the Milky Way Star Clusters catalog
presented in [19], which contains the coordinates, distances,
proper motions, and radial velocities for open and globular
clusters in the Galaxy. Complete information is available for 133
GCs, making it possible to compute their positions and their
$U,V,W$ space velocities in order to derive their Galactic orbits.

After the appearance of the catalog [19], more accurate proper
motions for a number of GCs were subsequently published. Thus, We
used the data of [20] for the clusters Terzan~1, Terzan~2, Terzan
4, Terzan 9, NGC 6522, NGC 6540, NGC 6558, NGC~6652, NGC~6681, and
Palomar~6, taking into account the corrections of [21]. The
absolute proper motions for these GCs were derived from an
analysis of ground-based observations using telescopes of the
European Southern Observatory, as well as observations with the
Hubble Space Telescope. Detailed studies of their individual
orbits can be found in [21, 22].

We used the proper motions derived in [23] using data from the
Gaia TGAS (Tycho-Gaia Astrometric Solution) catalog for NGC 104,
NGC 5272, NGC 6121, NGC 6397, and NGC 6656.

We used the absolute proper-motion components and heliocentric
distance for $\omega$~Cen obtained in [24] using observations with
the Hubble Space Telescope:
 $\mu_\alpha\cos\delta=-3.238\pm0.028$ milliseconds per year (mas/year),
 $\mu_\delta=-6.716\pm0.043$ mas/year, and
 $r = 5.20\pm0.25$~kpc.

We took the parameters of the center of the Gould Belt from [25],
where they were computed from data for molecular-cloud complexes
using VLBI observations of radio stars and maser sources; the
position of this center is
$(x_0,y_0,z_0)=(-118,54,-12)\pm(15,11,8)$~pc, and its mean
velocity relative to the Sun is
$(u_0,v_0,w_0)=(-13.6,-15.0,-6.4)\pm(0.7,0.7,0.6)$ km/s.

\section{METHOD}
Since the method used in our study was based on analyzing the
orbits of objects (GCs, the Gould Belt), we will spend some time
describing our chosen model for the Galaxy. We believe that the
most realistic model available is our refinement [26, 27] of the
Navarro–Frenk–White model [28], based on the most recent data,
supplemented with terms taking into account the influence of the
central bar and spiral-density wave.

 \subsection{Model for the Axially Symmetric Galactic Potential}
We present a model for the axially symmetric gravitational
potential of the Galaxy as a sum of three components: the central
spherical bulge, $\Phi_b(r(R,Z))$, the disk $\Phi_d(r(R,Z))$, and
the massive, spherical dark-matter halo $\Phi_h(r(R,Z))$:
 \begin{equation}
 \begin{array}{lll}
  \Phi(R,Z)=\Phi_b(r(R,Z))+\Phi_d(r(R,Z))+\Phi_h(r(R,Z)).
 \label{pot}
 \end{array}
 \end{equation}
Here, we used a cylindrical coordinate system ($R,\psi,Z$) with
its origin at the Galactic center. In Cartesian coordinates
$(X,Y,Z)$ with their origin at the Galactic center, the distance
to a star (the spherical radius) is $r^2=X^2+Y^2+Z^2=R^2+Z^2,$
where the $X$ axis is directed from the Sun toward the Galactic
center, the $Y$ axis is perpendicular to the $X$ axis and points
in the direction of the Galactic rotation, and the $Z$ axis is
perpendicular to the Galactic $(XY)$ plane and points in the
direction of the North Galactic pole. The gravitational potential
is expressed in units of 100 km$^2$ s$^{-2}$, distances in kpc,
masses in units of the mass of the Galaxy, $M_{gal}=2.325\times
10^7 M_\odot$, and the gravitational constant is taken to be
$G=1.$

The potentials of the bulge $\Phi_b(r(R,Z))$ and disk
$\Phi_d(r(R,Z))$ were taken to have the form proposed by Miyamoto
and Nagai [29]:
 \begin{equation}
  \Phi_b(r)=-\frac{M_b}{(r^2+b_b^2)^{1/2}},
  \label{bulge}
 \end{equation}
 \begin{equation}
 \Phi_d(R,Z)=-\frac{M_d}{\Biggl[R^2+\Bigl(a_d+\sqrt{Z^2+b_d^2}\Bigr)^2\Biggr]^{1/2}},
 \label{disk}
\end{equation}
where $M_b$ and $M_d$ are the masses of the corresponding
components and $b_b, a_d,$ and $b_d$ are scale parameters of the
components in kpc. The halo component was taken in accordance with
[28]:
 \begin{equation}
  \Phi_h(r)=-\frac{M_h}{r} \ln {\Biggl(1+\frac{r}{a_h}\Biggr)}.
 \label{halo-III}
 \end{equation}
Table 1 presents the parameters of the model for the Galactic
potential (2)--(4) from [26, 27], computed using the rotational
velocities of Galactic objects at distances $R$ out to $\sim$200
kpc. When deriving the corresponding Galactic rotation curves, we
used $R_\odot=8.3$ kpc for the Galactocentric distance and
$V_\odot=244$ km/s for the linear velocity of the local standard
of rest around the center of the Galaxy.

 {\begin{table}[t]                                    
 \caption[]
 {\small\baselineskip=1.0ex
 Parameters of the Galactic potential model, $M_0=2.325\times10^7 M_\odot$
  }
 \label{t:model-III}
 \begin{center}\begin{tabular}{|c|c|r|}\hline
 Parameter    &     Value         \\\hline
 $M_b$($M_0$) &    443$\pm27$     \\
 $M_d$($M_0$) &   2798$\pm84$     \\
 $M_h$($M_0$) &  12474$\pm3289$   \\
 $b_b$(kpc)   & 0.2672$\pm0.0090$ \\
 $a_d$(kpc)   &   4.40$\pm0.73$   \\
 $b_d$(kpc)   & 0.3084$\pm0.0050$ \\
 $a_h$(kpc)   &    7.7$\pm2.1$    \\\hline
 \end{tabular}\end{center}\end{table}}

 \subsection{The Bar and Spiral Structure of the Galaxy}
We selected a triaxial ellipsoid model in accordance with [30] for
the potential of the bar:
\begin{equation}
  \Phi_{bar} = -\frac{M_{bar}}{(q_b^2+X^2+[Ya_b/b_b]^2+[Za_b/c_b]^2)^{1/2}},
\label{bar}
\end{equation}
where $X=R\cos\vartheta, Y=R\sin\vartheta$, $a_b, b_b, c_b$ are
the three semi-axes of the bar; $q_b$ is the length of the bar;
$\vartheta=\theta-\Omega_{bar}t-\theta_{bar}$, $tg(\theta)=Y/X$,
$\Omega_{bar}$ is the angular speed of the bar; $t$ is the
integration time; and $\theta_{bar}$ is the inclination of the bar
relative to the $X$ and $Y$ axes, measured from the line joining
the Sun and the Galactic center (the $X$ axis) to the major axis
of the bar in the direction of the Galactic rotation. We adopted
the angular speed of the bar $\Omega_{bar}=55$ km s$^{-1}$
kpc$^{-1}$, in accordance with the estimate of [31].

When the spiral-density wave is taken into account [32, 33], the
right-hand side of formula (1) is supplemented with the term [34]:
\begin{equation}
 \Phi_{sp} (R,\theta,t)= A\cos[m(\Omega_p t-\theta)+\chi(R)],
 \label{Potent-spir}
\end{equation}
where
 $$
 A= \frac{(R_0\Omega_0)^2 f_{r0} \tan i}{m},$$$$
 \chi(R)=- \frac{m}{\tan i} \ln\biggl(\frac{R}{R_0}\biggr)+\chi_\odot,
 $$
Here, $A$ is the amplitude of the spiral-wave potential, $f_{r0}$
the ratio of the radial component of the perturbation to the total
gravitation of the Galaxy, $\Omega_p$ the angular velocity of the
wave's rigid-body rotation, $m$ the number of spiral arms, $i$ the
pitch angle of the arms ($i<0$ for a trailing pattern), $\chi$ the
phase of the radial wave ($\chi=0^\circ$ corresponds to the center
of the arm), and $\chi_\odot$ the Sun's radial phase in the spiral
wave. We adopted the following parameters for the spiral wave:
 \begin{equation}
 \begin{array}{lll}
 m=4,\\
 i=-13^\circ,\\
 f_{r0}=0.05,\\
 \chi_\odot=-120^\circ,\\
 \Omega_p=20~\hbox {km s$^{-1}$ kpc$^{-1}$}.
 \label{param-spiral}
 \end{array}
 \end{equation}

 \subsection{Equations of Motion}
The equations of motion of a test particle in the Galactic
potential have the form
\begin{equation}
 \begin{array}{llllll}
 \dot{X}=p_X, ~~\dot{Y}=p_Y, ~~\dot{Z}=p_Z,\\
 \dot{p}_X=-\partial\Phi/\partial X,\\
 \dot{p}_Y=-\partial\Phi/\partial Y,\\
 \dot{p}_Z=-\partial\Phi/\partial Z,
 \label{eq-motion}
 \end{array}
\end{equation}
where $p_X, p_Y,$ and $p_Z$ are canonical momenta, and a dot
denotes a derivative with respect to time. We integrated Eqs. (8)
using a fourth-order Runge–Kutta algorithm.

We took the peculiar velocity of the Sun relative to the Local
Standard of Rest to be
$(u_\odot,v_\odot,w_\odot)=(11.1,12.2,7.3)$~km/s, as was
determined by Sch\"onrich et al. [35]. Here, heliocentric
velocities correspond to a set of moving Cartesian coordinates,
with $u$ directed towards the Galactic center, $v$ in the
direction of the Galactic rotation, and $w$ perpendicular to the
Galactic plane, towards the north Galactic pole.

Let the initial positions and space velocities of a test particle
in the heliocentric coordinate system be
$(x_0,y_0,z_0,u_0,v_0,w_0)$. The initial positions and velocities
of the test particle in Galactic Cartesian coordinates are then
given by
\begin{equation}
 \begin{array}{llllll}
 X=R_0-x_0, ~~Y=y_0, ~~Z=z_0+h_\odot,\\
 U=-(u_0+u_\odot),\\
 V=v_0+v_\odot+V_0,\\
 W=w_0+w_\odot,
 \label{init}
 \end{array}
\end{equation}
where $h_\odot=16$~pc is the height of the Sun above the Galactic
plane.

 \subsection{Statistical Modeling}
We performed Monte Carlo statistical modeling for each GC. In
these simulations, we added random errors to the object's
coordinates $(X,Y,Z)$ and space velocities $(U,V,W),$ which were
computed, together with their uncertainties, taking into account
the random errors in the distances, proper motions, and radial
velocities.

Several methods for estimating distances to GCs are known: RR
Lyrae variable stars, dynamical methods, eclipsing variables,
Cepheids, fitting suitable isochrones, etc. The catalog [36] gives
an extensive bibliography concerning distance-determination
techniques for each GC. An analysis of the data in this catalog
shows that each of the methods has a relative error of $\sim$10\%.
Francis and Anderson [36] calculated the mean distances of each of
154 Galactic GCs and estimated the random errors of the derived
distances to be in the range 1--5\%.

Thus, in our simulations, we adopted a relative random error in
the distance of 5\% for all the GCs. The confidence intervals for
the derived points where the GCs cross the Galactic plane, as well
as for the model orbits in the Gould Belt, are estimated at the
99.7\% $(3\sigma)$ probability level.

 \subsection{Time Characteristics of Star Formation}
It is clear that some time must elapse after the impact of a GC
onto the Galactic plane before stars will be formed. Following
[37], we based our study on the relation
\begin{equation}
 t=t_{\rm C}+t_{\rm SF}+t_{\rm A},
 \label{time-3}
\end{equation}
where $t$ is the time elapsed from the crossing of the Galactic
disk by the GC to the present time, $t_{\rm C}$ the time between
the crossing and the onset of star formation, $t_{\rm SF}$ the
duration of the star formation, and $t_{\rm A}$ the age of the
structure formed (in particular, the Gould Belt).

The value of the first term in Eq. (10) is known only with a large
uncertainty, and is in the range 0--30 million years. For example,
$t_{\rm C}=15$~million years according to the estimate of [14]
obtained from simulations of an impact of a high-velocity cloud
onto the disk. According to Wallin et al. [38], this time interval
is $t_{\rm C}=30$~million years. In the model computations of
Bekki [18], the time interval for star formation is in the range
$t_{\rm C}=7-15$~million years. According to [39], the second term
is $t_{\rm SF}=0.2$ million years (for a stellar mass
$M>M_\odot);$ since this is small compared to the other terms, we
can neglect it in a first rough estimate. Finally, we adopted an
age for the Gould Belt of $t_{\rm A}=60$ million years.

\begin{figure}[t]
{\begin{center}
   \includegraphics[width=0.99\textwidth]{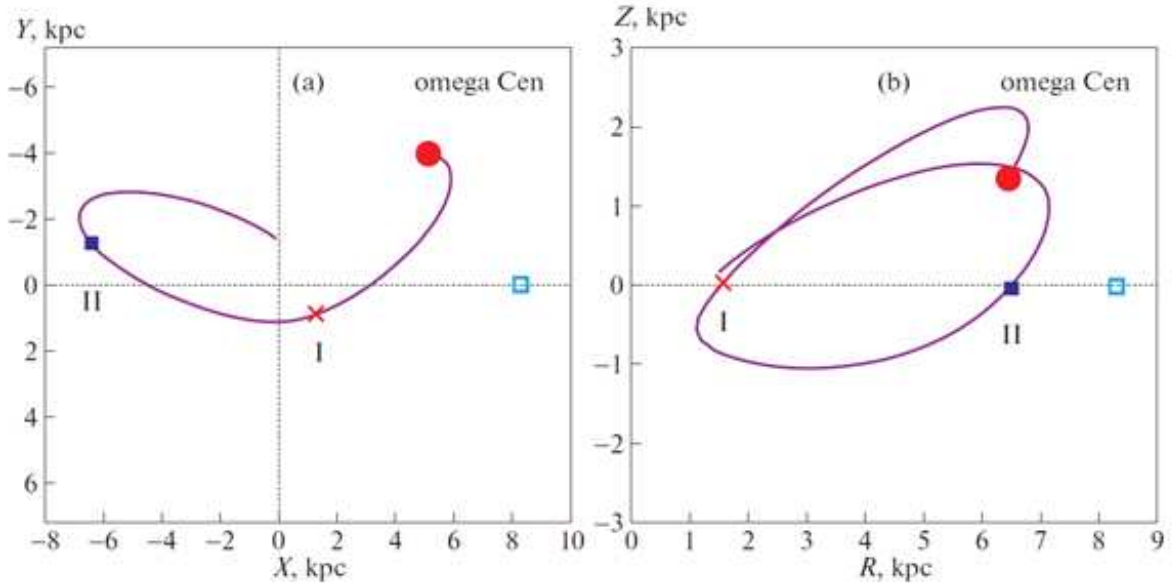}
 \caption{\small
Projection of the orbit of $\omega$~Cen onto (a) the Galactic
plane $XY$ and (b) the accompanying plane $RZ.$ The orbit was
computed 130 million years into the past using the axially
symmetric potential model. The center of the Galaxy is at the
coordinate origin, the open (light blue) squares show the position
of the Sun, the x's marked I show the times of the last
Galactic-plane crossing, and the filled (blue) squares marked II
show the times of the next-to-last Galactic-plane crossing. The
current position of $\omega$~Cen is plotted by the large filled
(red) circles.
  } \label{f1}
\end{center}}
\end{figure}
\begin{figure}[t]
{\begin{center}
   \includegraphics[width=0.99\textwidth]{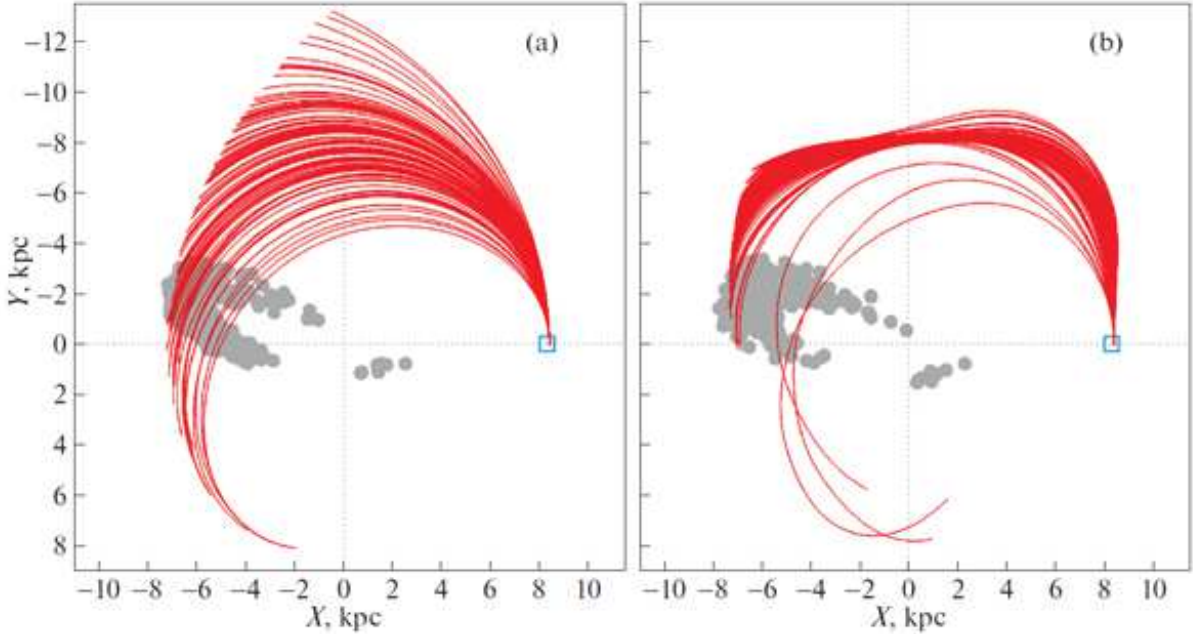}
 \caption{\small
Confidence intervals for the crossing of the Galactic plane $XY$
by $\omega$~Cen (gray circles). The colored (red) curves show
model trajectories for the center of the Gould Belt derived using
the Monte Carlo method, (a) based on the axially symmetric
potential model for an interval of 80 million years into the past
and (b) taking into account the influence of the bar and spiral
density wave for an interval of 96 million years into the past.
The Galactic center is at the coordinate origin; the square marks
the position of the Sun.
  } \label{f2}
\end{center}}
\end{figure}

 \section{RESULTS AND DISCUSSION}
Let us first consider Fig. 1, which shows projections of the orbit
of the GC $\omega$~Cen onto the Galactic plane $XY$ and onto the
accompanying plane $RZ.$ The orbit was computed 130 million years
into the past using an axially symmetric potential model. The
times of crossing of the Galactic plane are marked in the figure.
The first crossing (I) occurred 49 million years ago, when the GC
passed from the southern to the northern Galactic hemisphere (a
S--N transition). The second crossing (II) was 80 million years
ago, when the GC passed from the northern to the southern
hemisphere (a N--S transition).

Note that the direction of the passage is not the defining factor
in the formation of the Gould Belt. However, since the formation
process is not instantaneous, and the initial distribution of the
gas is not uniform, we expect the presence of an age gradients for
individual components of the Gould Belt along the trajectory of
the impacting body. For example, the mean age of the main members
of the Scorpius-Centaurus association (the northern part of the
Gould Belt) is 10--20 million years greater than the mean age of
members of the Orion association (the southern part of the Gould
Belt). At the same time, the region of maximum density with the
highest star-formation rate should be located at the center of the
Gould Belt, as was demonstrated in [18]. By the way, the impacting
body crosses the Galactic plane from the South to the North in
this model.

Figure 2 displays the confidence interval for the points where
$\omega$~Cen crosses the Galactic plane $XY$ and the trajectories
of the center of the Gould Belt obtained using the Monte Carlo
method. The orbits are plotted for both the axially symmetric
potential model and a model including the influence of the bar and
spiral-density wave. Since the times of a GC's crossing of the
Galactic plane are somewhat different for different potentials, we
integrated the model orbits of the center of the Gould Belt for
different appropriate time intervals. The distribution of points
where $\omega$~Cen crosses the Galactic plane corresponds to
crossing II. We estimated the probability $p$ from the number of
Gould Belt trajectories that intersect the distribution of
crossing points for the time $t.$ The panels of Fig. 2 each show
100 model trajectories for the Gould Belt. Of these, 15 (Fig. 2a)
and 20 (Fig. 2b) trajectories intersect the area where the
crossing points are distributed; the corresponding probabilities
are $p=15\%$ and $p=20\%,$ respectively.

Table 2 presents the nominal characteristics for the approach of
the center of the Gould Belt to the site of the next-to-last
crossing of the Galactic plane by $\omega$~Cen. The orbits and
their characteristics were computed using three models for the
potential: (a) an axially symmetric model in accordance with
(2)--(4), (b) the axially symmetric model with an added
contribution from the bar (5), and (c) the axially symmetric model
with added contributions from the bar and spiral density wave (6).
These three options for the potential are indicated in the first
line of Table 2. For each crossing time $t,$ we computed the
distance between the position of the Gould Belt and the site where
a GC crossed the Galactic plane, $\Delta r_t=\sqrt{\Delta
X^2+\Delta Y^2+\Delta Z^2},$ where $\Delta X, \Delta Y,$ and
$\Delta Z$ are the differences between the coordinates of the
Gould Belt and the GC. The $\Delta r_t$ value in the second column
of Table 2 was obtained by averaging all the model orbits. The
crossing time $t$ (also corresponding to an average of all the
model orbits) is given in the third column, and the fourth column
presents the angle $\gamma$ at which the GC crosses the Galactic
plane (the angle between the orbital plane of the GC and the $XY$
plane). As was noted in the Introduction, the more acute the angle
$\gamma$, the more effective the compression of the Galactic disk.
The last column of Table 2 contains the probability $p$ for the
Gould Belt trajectories to intersect the distribution of the
crossing points.

It follows from Fig. 2 and a comparison of the data in Table 2
that the influence of the spiral-density wave is a significant
factor. Including this influence leads to a considerably lower
scatter of the model orbits, and the approach parameter $\Delta
r_t$ is smaller than in the axially symmetric case.

 \begin{table}[t]
 \caption[]
  {\small
Parameters of the approach of the center of the Gould Belt to the
site of the next-to-last Galactic-plane crossing by $\omega$~Cen
  }
  \begin{center}  \label{t:02}
  \small
  \begin{tabular}{|l|c|c|c|c|c|c|}\hline
 Potential model            & $\Delta r_t,$ & $t,$ million & $\gamma, ^\circ$ & $p,$ \% \\
                            &       kpc     &    years ago &                  &       \\\hline
 a) axially symmetric case  &           4.7 &         $80$ & $34\pm8$ & 15 \\
 b) + bar                   &           4.8 &         $81$ & $35\pm8$ & 16 \\
 c) + bar + spiral wave     &           2.5 &         $96$ & $40\pm8$ & 20 \\\hline

 \end{tabular}\end{center}
 \end{table}
 \begin{table}[t]
 \caption[]
  {\small
Globular clusters whose crossings of the Galactic plane were near
the center of the Gould Belt about 100 million years ago
  }
  \begin{center}  \label{t:03}
  \small
  \begin{tabular}{|c|r|c|c|c|c|c|c|c|c}\hline
   Globular    & $n_c$ & $\Delta r_t,$ & $t,$ million & $M/M_\odot$ &   Source & Crossing & $\gamma, ^\circ$& $p,$ \% \\
    cluster    &       &       kpc     &    years ago &             &          &          &                          & \\\hline
  $\omega$~Cen &  II   &           2.5 & $96$ & $4.00\times10^6$ & [40]  & N--S & $40\pm8$ & 20 \\
      NGC 7078 &  II   &           1.0 & $98$ & $0.65\times10^6$ & [41]  & S--N & $39\pm7$ & 74 \\
      NGC 6341 &  II   &           1.8 & $82$ & $0.30\times10^6$ & [41]  & N--S & $57\pm9$ & 72 \\
      NGC 6838 & III   &           3.7 & $86$ & $0.02\times10^6$ & [42] &  --- & ---  & --- \\
       NGC 104 &   I   &           7.7 & $53$ & $0.84\times10^6$ & [39] &  --- & ---  & --- \\
      NGC 6760 &  II   &           8.3 & $52$ & $0.25\times10^6$ & [43] &  --- & ---  & --- \\
      NGC 6749 & III   &           9.4 & $52$ &    ---           &  --- &  --- & ---  & --- \\\hline
 \end{tabular}\end{center}
 \end{table}
\begin{figure}[t]
{\begin{center}
   \includegraphics[width=0.99\textwidth]{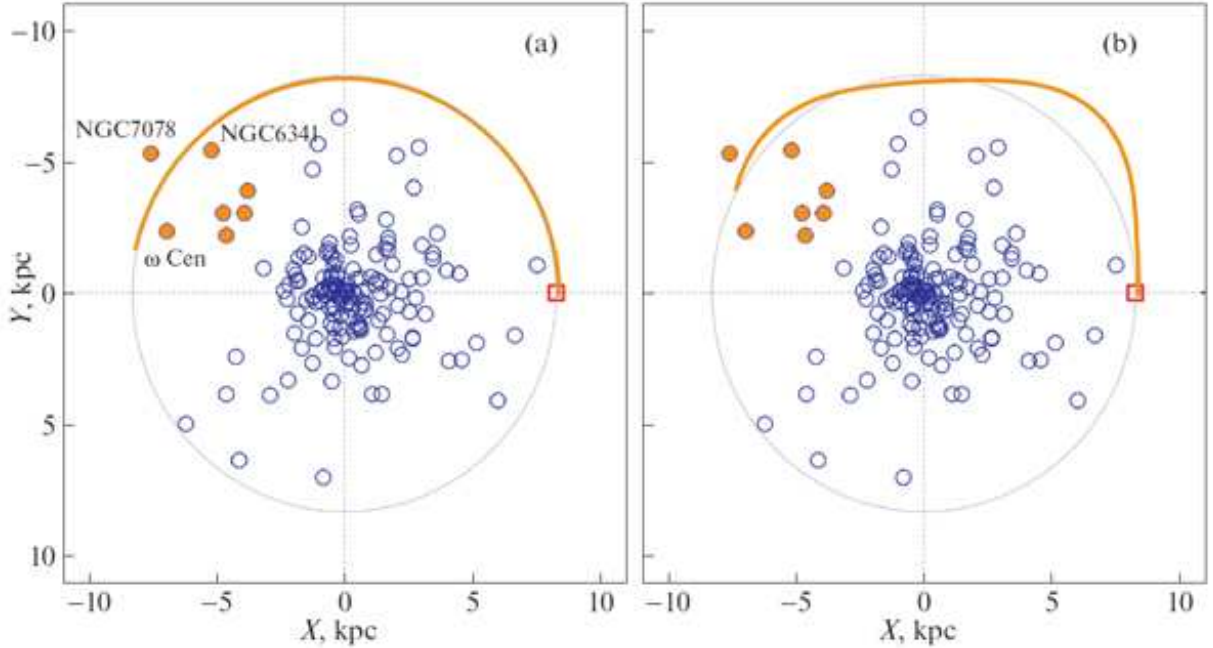}
 \caption{\small
Model points where GCs cross the Galactic plane $XY$ (blue
circles) and model trajectories for the center of the Gould Belt.
Curves are plotted 100 million years into the past (a) using the
axially symmetric potential model and (b) including with the
influence of the bar and the spiral-density wave. The center of
the Galaxy is at the coordinate origin, the (red) squares show the
position of the Sun, and the filled (yellow) circles show the
positions of known clusters.
  } \label{f3}
\end{center}}
\end{figure}
\begin{figure}[t]
{\begin{center}
   \includegraphics[width=0.99\textwidth]{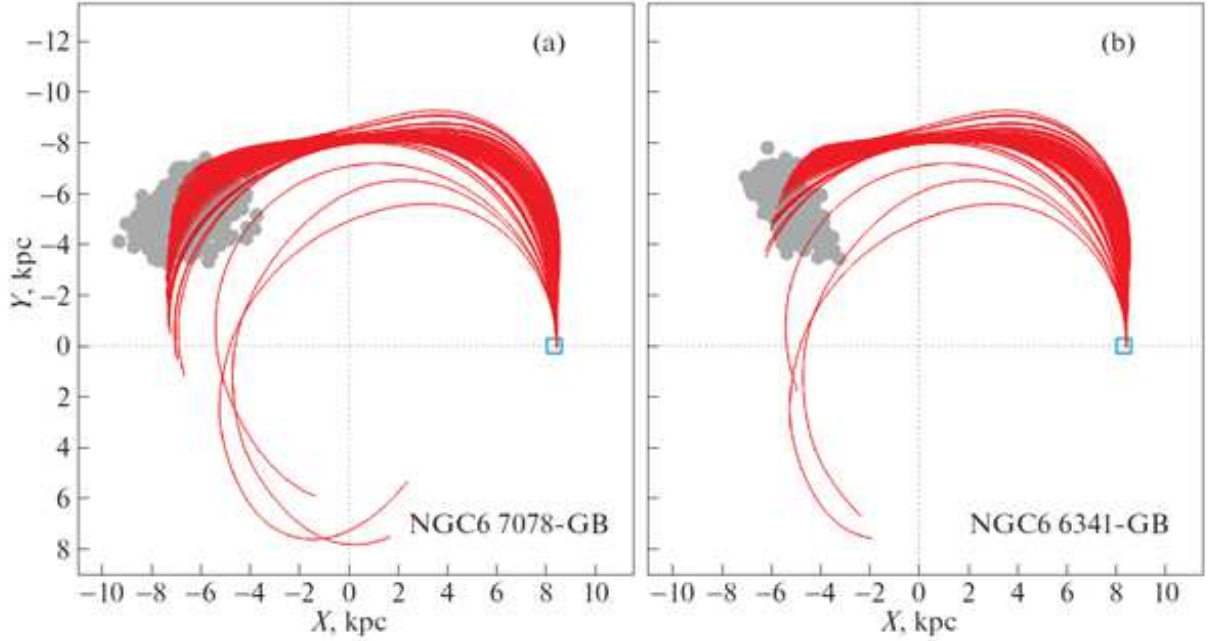}
 \caption{\small
(a) Confidence intervals for the points where NGC 7078 crosses the
Galactic plane $XY$ (gray circles) and model trajectories for the
center of the Gould Belt (red curves) plotted 98 million years
into the past. (b) Same as (a) for NGC 6341 (gray circles),
plotting the model trajectories for the center of the Gould Belt
(red curves) 82 million years into the past. The center of the
Galaxy is at the coordinate origin, and the (blue) squares mark
the position of the Sun.
  } \label{f4}
\end{center}}
\end{figure}
\begin{figure}[t]
{\begin{center}
   \includegraphics[width=0.8\textwidth]{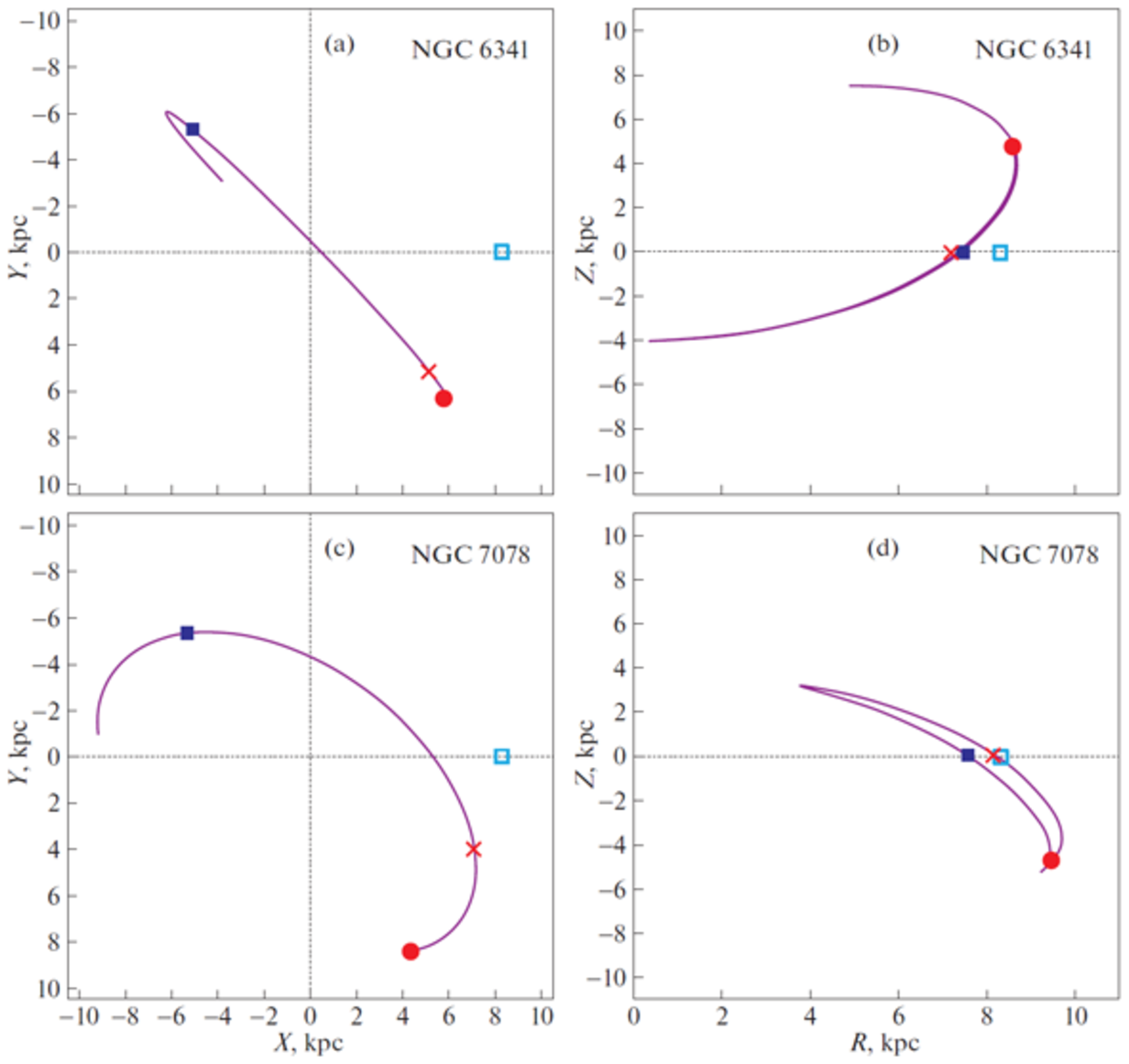}
 \caption{\small
Orbit of NGC 6341 projected onto (a) the Galactic plane $XY$ and
(b) the accompanying plane $RZ.$ The same is shown for NGC 7078 in
panels (c) and (d), respectively. The models were computed 150
million years into the past using the axially symmetric potential
model. The (red) crosses show the times of the last crossings of
the Galactic plane, and the solid (blue) squares the times of the
next-to-last crossings of the Galactic plane. The current position
of the GC is plotted as a red circle, the center of the Galaxy is
at the coordinate origin, and the (blue) squares mark the position
of the Sun.
  } \label{f5}
\end{center}}
\end{figure}

Figure 3 shows the model crossing points of the Galactic plane
$XY$ by GCs and model trajectories of the center of the Gould Belt
plotted using the axially symmetric potential model (Fig. 3a) and
the potential model taking into account the influence of the bar
and spiral-density wave (Fig. 3b). We selected GCs with crossing
times in the range 50--100 million years for this figure. We
decided not to plot points with crossing times younger than 50
million years to avoid crowding the figure, particularly as there
are no points close to the Gould Belt trajectory among them. The
filled (yellow) circles mark seven candidate GCs undergoing close
approaches with the Gould Belt orbit; names are indicated for the
three GCs closest to this orbit. We selected these seven
candidates based on the proximity of their positions to the Gould
Belt orbit.

Table 3 presents the parameters for these seven candidate GCs that
could have undergone a close approach with the Gould Belt orbit in
the past. All the data in Table 3 were obtained using the most
complete model for the Galactic potential (including the bar and
spiral-density wave). The columns of Table 3 give (1) the names of
the GCs, (2) the ordinal number nc for a GC's crossing of the
Galactic plane in the past, (3) the distances $\Delta r_t$, (4)
the time intervals $t,$ (5)--(6) mass estimates for the GCs and
corresponding references, (7) the type of crossing, (8) the
crossing angle $\gamma,$ and (9) the probabilities $p$ for the
Gould Belt trajectories to intersect the distribution of crossing
points.

Table 3 shows that the mass of $\omega$~Cen is an order of
magnitude higher than the masses of the other GCs in this table.
We could not find a mass estimate for NGC 6749 in the literature.
However, it is known to have a fairly diffuse appearance, and
Rosino et al. [45] remark that it could even be an open cluster.
Thus, we expect the mass of NGC 6749 not to be very high.

The mass of NGC 6838 is too low for it to be the body whose impact
formed the Gould Belt. We can also exclude the GCs in the last
three lines of the table, namely NGC 104, NGC 6760, and NGC 6749,
from the list of candidates for the following reason. All three of
these GCs crossed the Galactic disk some 52 million years ago, at
which time the distance between each of them and the Gould Belt
orbit was too large to influence the possible formation of the
Gould Belt (the distances $\Delta r_t$ exceeded 90\% of the
heliocentric distance to the site where each GC crossed the
Galactic plane).

Figure 3 also contains a point with $(X,Y)\approx(0,-7)$~kpc that
is fairly close to the Gould Belt trajectory. This is the site
where the Galactic plane was crossed by Palomar 10 at time
$-99$~million years. This GC is of no interest for us, as its
$\Delta r_t$ was too large at that time. The situation is similar
for six other objects located near the area with coordinates
$(X,Y)\approx(0,-5)$~kpc: their distances to the Gould Belt orbit
were too large at their crossing times ($\Delta r_t>10$~kpc).

Figure 4a shows the confidence intervals for the points $XY$ where
the Galactic plane was crossed by NGC 7078 and the model
trajectories for the center of the Gould Belt plotted 98 million
years into the past, and Fig. 4b shows the same for NGC 6341, with
the model Gould Belt trajectories plotted 82 million years into
the past. Here, we used the full model for the potential,
including the influence of the bar and spiral density wave.

Comparing Fig. 4 and Fig 2b, we can see that NGC 7078 and NGC 6341
have larger fractional intersection zones between the terminal
points of the Gould Belt orbits and the distribution of the GC
crossing points than does $\omega$~Cen. Thus, the probability $p$
of a close approach to the Gould Belt orbit is higher for NGC 7078
and NGC 6341 than for $\omega$~Cen.

Note that, in the case of $\omega$~Cen, NGC 7078, and NGC 6341,
the formation of the Gould Belt requires a time that is
approximately double the time scale indicated by the numerical
experiments of Bekki [18]. The time interval $t$ in (10) in his
computations was about 45 million years, with $t_A=30$~million
years.

The Galactic orbits of $\omega$~Cen, NGC 6341, and NGC 7078 all
differ considerably from circles. The orbit of $\omega$~Cen is
shown in Fig. 1, while Fig. 5 shows the orbits of NGC 6341 and NGC
7078 computed 150 million years into the past using the axially
symmetric potential model. It is of interest to obtain some idea
of the velocity $V_{tot}$ of the GCs relative to the medium
through which they were moving at the time of their crossing the
Galactic plane about 100 million years ago. To address this, we
first computed their cylindrical velocities $V_R,V_\theta,V_Z$ and
the crossing time II, which yielded
 $(V_R,V_\theta,V_Z)=(-95,-68,-85)$~km/s for $\omega$~Cen,
 $(-130,14,-209)$~km/s for NGC 6341, and
 $(121,157,165)$~km/s for NGC 7078.

Note that $\omega$~Cen has a retrograde orbit, and lags behind the
Galactic rotation by $\Delta V_\theta\approx68$~km/s. Gas clouds
in which stars can potentially form move along essentially
circular orbits, in accordance with the Galaxy’s rotation curve;
their velocity at the solar distance $R_0$ is
$(V_\theta)_{R=R_0}=V_0=244$~km/s, the value we used in our
potential. We see from Fig. 3 that the crossing points for all
three candidates were near the solar circle about 100 million
years ago, so that we can use the velocity $V_0$ as the circular
velocity of the medium in rough estimates. We then obtain for the
vector velocity of $\omega$~Cen relative to the gaseous medium
$(V_R,V_0-V_\theta,V_Z)=(-95,312,-85)$~km/s, with the absolute
value of the total velocity being $V_{tot}=337$~km/s.

NGC 6341 essentially does not participate in the Galactic
rotation, but has considerable radial and vertical velocities.
These properties are clearly seen in Fig.~5a, where the GC moves
essentially through the center of the Galaxy, as well as in
Fig.~5b, where NGC 6341 ascends to $Z_{max}=10.5$~kpc above the
Galactic plane. Using the approach described above, we found
$V_{tot}=337$~km/s.

NGC 7078 rotates with $V_\theta=157$~km/s, and also has
considerable radial and vertical velocities, as is shown by Figs.
5c, d. For this cluster, $V_{tot}=222$~km/s.

Thus, we can conclude that NGC 6341 and NGC 7078 possessed
comparable energies in their collision with the medium; the small
difference in their masses was compensated by the difference in
their velocities. However, there is no doubt that $\omega$~Cen is
leader in terms of its impact energy.

\section*{CONCLUSIONS}
We have computed Galactic orbits for 133 globular clusters in our
Galaxy using modern measurements of their proper motions, radial
velocities, and distances. For each GC, we identified points of
past crossings of the Galactic plane $XY,$ considering not only
the last crossing, but also several earlier crossings (to the
fifth last).

Our analysis of the distribution of these points shows three of
the GCs to be located very close to the trajectory of the center
of the Gould Belt at times from --80 to --100 million years. These
GCs are NGC 7078, NGC 6341, and $\omega$~Cen. The distance between
each of these and the Gould Belt orbit at the crossing time is
$\Delta r_t<2.5$~kpc (i.e., $<20\%$ of their heliocentric distance
at the crossing time $t),$ within the random uncertainties in the
distances to the Galactic orbits.

We integrated the orbits of the GCs and the of center of the Gould
Belt using an axially symmetric model for the potential, and also
taking into account the contributions of the central bar and
spiral-density wave. We found that the difference in the approach
times when the spiral-density wave is taken into account can reach
10--15 million years.

$\omega$~Cen, whose mass, $4\times10^6 M_\odot$, is higher than
the masses of the other candidates by an order of magnitude, is of
primary interest. At the same time, the sites of past crossings of
the Galactic plane by NGC 7078 and NGC 6341 are closer to the
trajectory of the Gould Belt than the site for $\omega$~Cen. The
time for the close approach to the Gould Belt that is closest to
the present time is $t=82$ million years ago for NGC 6341, while
the time that is farthest is that for NGC 7078, $t=98$ million
years ago. In our opinion, each of these three clusters---NGC
7078, NGC 6341, and $\omega$~Cen---could be a candidate for the
body whose passage through the Galactic disk may have provoked the
formation of the Gould Belt.

The analysis of different sources in the literature indicates a
fairly good agreement in the derived mean radial velocities and
distances to the GCs studied. On the other hand, their measured
proper motions have large uncertainties. It is important to use
the most reliable proper motions possible for objects with large
heliocentric distances. Thus, it will be important to confirm our
results after the Gaia mission [46] is completed.

 \subsubsection*{ACKNOWLEDGEMENTS}
The authors thank the referee for helpful remarks that have
enabled us to improve this paper. Our study was supported by Basic
Research Program P--28 of the Presidium of the Russian Academy of
Sciences (subprogram ``Space: Studies of Basic Processes and their
Interrelations'').

 \bigskip\subsubsection*{REFERENCES}

 {\small
 \quad ~1. J.A. Frogel and R. Stothers, Astron. J. 82, 890 (1977).

 2. Yu. N. Efremov, Star Formation Centers in Galaxies (Nauka, Moscow, 1989) [in Russian].

 3. W.G.L. P\"oppel, Fundam. Cosmic Phys. 18, 1 (1997).

 4. P.T. de Zeeuw, R. Hoogerwerf, J.H.J. de Bruijne, et al., Astron. J. 117, 354 (1999).

 5. J. Torra, D. Fern\'andez, and F. Figueras, Astron. Astrophys. 359, 82 (2000).

 6. V.V. Bobylev, Astrophysics 57, 625 (2014).

 7. A. Blaauw, Koninkl. Ned. Akad. Wetenschap. 74 (4) (1965).

 8. C.A. Olano, Astron. J. 112, 195 (1982).

 9. P.O. Lindblad, Astron. Astrophys. 363, 154 (2000).

 10. V.V. Bobylev, Astron. Lett. 30, 785 (2004).

 11. V.V. Bobylev, Astron. Lett. 32, 816 (2006).

 12. C.A. Olano, Astron. J. 121, 295 (2001).

 13. C.A. Perrot and I. A. Grenier, Astron. Astrophys. 404, 519 (2003).

 14. J.R. L\'epine and G. Duvert, Astron. Astrophys. 286, 60 (1994).

 15. F. Comer\'on and J. Torra, Astron. Astrophys. 261, 94 (1992).

 16. F. Comer\'on and J. Torra, Astron. Astrophys. 281, 35 (1994).

 17. V.V. Levy, Astron. Astrophys. Trans. 18, 621 (2000).

 18. K. Bekki, Mon. Not. R. Astron. Soc. 398, L36 (2009).

 19. N.V. Kharchenko, A.E. Piskunov, S. R\"oser, et al., Astron. Astrophys. 558, A53 (2013).

 20. L.J. Rossi, S. Ortolani, B. Barbuy, et al.,
     Mon. Not. R. Astron. Soc. 450, 3270 (2015).

 21. A. P\'erez-Villegas, L. Rossi, S. Ortolani, et al.,
     Publ. Astron. Soc. Australia 35, 21 (2018).

 22. V.V. Bobylev and A.T. Bajkova, Astron. Rep. 61, 551 (2017).

 23. L.L. Watkins and R.P. van der Marel, Astrophys. J. 839, 89 (2017).

 24. M. Libralato, A. Bellini, L. R. Bedin, et al., Astrophys. J. 854, 45 (2018).

 25. V.V. Bobylev, Astron. Lett. 42, 544 (2016).

 26. A.T. Bajkova and V.V. Bobylev, Astron. Lett. 42, 567 (2016).

 27. A.T. Bajkova and V.V. Bobylev, Open Astron. 26, 72 (2017).

 28. J.F. Navarro, C.S. Frenk, and S.D.M. White, Astrophys. J. 490, 493 (1997).

 29. M. Miyamoto and R. Nagai, Publ. Astron. Soc. Jpn. 27, 533 (1975).

 30. J. Palou\v{s}, B. Jungwiert, and J. Kopeck\'y, Astron. Astrophys. 274, 189 (1993).

 31. V.V. Bobylev and A.T. Bajkova, Astron. Lett. 42, 228 (2016).

 32. C.C. Lin and F.H. Shu, Astrophys. J. 140, 646 (1964).

 33. C.C. Lin, C. Yuan, and F.H. Shu, Astrophys. J. 155, 721 (1969).

 34. D. Fernandez, F. Figueras, and J. Torra, Astron. Astrophys. 480, 735 (2008).

 35. R. Sch\"onrich, J. Binney, and W. Dehnen, Mon. Not. R. Astron. Soc. 403, 1829 (2010).

 36. C. Francis and E. Anderson, Mon. Not. R. Astron. Soc. 441, 1105 (2014).

 37. D. Vande Putte and M. Cropper, Mon. Not. R. Astron. Soc. 392, 113 (2009).

 38. J.F. Wallin, J.L. Higdon, and L. Staveley-Smith, Astrophys. J. 459, 555 (1996).

 39. C.F. McKee and J.C. Tan, Nature (London, U.K.) 416, 59 (2002).

 40. H. Nakaya, M. Watanabe, M. Ando, T. Nagata, and S. A. Sato, Astron. J. 122, 876 (2001).

 41. A. Sollima and H. Baumgardt, Mon. Not. R. Astron. Soc. 471, 3668 (2017).

 42. B. Kimmig, A. Seth, I.I. Ivans, et al., Astron. J. 149, 53 (2015).

 43. A. Bellini, P. Bianchini, A.L. Varri, et al., Astrophys. J. 844, 167 (2017).

 44. D.E. McLaughlin and R.P. van Der Marel, Astrophys. J. Suppl. 161, 304 (2005).

 45. L. Rosino, S. Ortolani, B. Barbuy, and E. Bica, Mon. Not. R. Astron. Soc. 289, 745 (1997).

 46. T. Prusti, J.H.J. de Bruijne, A.G.A. Brown, et al., Astron. Astrophys. 595, A1 (2016).

 }
 \end{document}